\documentclass[pra,twocolumn,aps,showpacs,superscriptaddress,groupedaddress]{revtex4-1}

\usepackage{hyperref}
\usepackage{graphicx}
\usepackage{amsmath}
\usepackage{amsfonts}
\usepackage{amssymb}
\usepackage{epsfig}
\usepackage{subfigure}
\usepackage[usenames,dvipsnames]{color}
\usepackage{bm}
\everymath{\displaystyle}
\newcommand{\bea}{\begin{eqnarray}}
\newcommand{\eea}{\end{eqnarray}}
\begin{document}

\title{Mode bifurcation in the Rayleigh-Taylor instability of binary 
       condensates}

\author{Arko Roy} 
\email{arkoroy@prl.res.in}
\author{S. Gautam} 
\author{D. Angom}
\affiliation{Physical Research Laboratory,
         Navarangpura, Ahmedabad - 380 009, Gujarat, India\\}

\date{\today}


\begin{abstract}

We examine the generation and subsequent evolution of Rayleigh Taylor
instability in anisotropic binary Bose-Einstein condensates. Considering a 
pancake-shaped geometry, to initiate the instability we tune 
the intraspecies interaction and analytically study the normal
modes of the interface in elliptic cylindrical coordinates. The normal modes
are then Mathieu functions and undergoes bifurcation at particular 
values of anisotropy and ratio of number of atoms. We find that the analytical 
estimates of the bifurcation parameters are in good agreement with the 
numerical results. 

\end{abstract}


\maketitle


\section {Introduction}
Rayleigh-Taylor instability (RTI)\cite{rayleigh_83,taylor_50,chandrasekhar_81} 
is the instability of an interface between  two fluids, which sets in, 
when a layer of lighter fluid supports a denser one or when a lighter fluid 
pushes a denser one, under the influence of gravitational field or some 
external potential. This occurs due to unfavourable energy conditions and as 
a result, the fluids tend to swap their positions. Any perturbation arising on 
the interface, however, small it may be, grows exponentially due to RTI and 
turbulent mixing of the fluids occur. During the process of mixing, the 
interface gets deformed and develops complicated 
non-linear patterns with mushroom shapes. The phenomenon of RTI is widely 
common in nature, ranging from convection of water to dusty plasma in 
atmosphere to supernova explosions\cite{sen_10,bychkov_06,cabot_06}. 
Recently, RTI has also been observed in a trapped two-species 
Bose-Einstein condensate (TBEC)\cite{sasaki_09}, where, intraspecies
scattering length plays a major role . Systems of trapped TBEC's that have been so far studied for observing RTI are a tight, symmetric 
pancake-shaped system in which the components separate out radially, 
a cigar-shaped trap in which phase-separation occurs in the axial direction 
and a perfectly spherical symmetric trap
\cite{gautam_10,kadokura-12,kobyakov_11}. Though experimental studies on
RTI are rare, theoretical studies on interfacial instabilities has been a
major research topic in the recent years. 
Other instabilities such as, Kelvin-Helmholtz instability 
(KHI)\cite{takeuchi_10}, Faraday 
instability have also been predicted in TBEC\cite{nicolin_07,aranya_2008}.
Experimental observation of quantum KHI and Faraday waves in BEC can be
found in Refs.\cite{blaauwgeers_02,engels_07}.\\
In the present work, we study RTI in a TBEC confined in a harmonic trapping
potential. The intraspecies and interspecies interaction between the atoms
are taken to repulsive. The initial state of the TBEC that we consider for
our study, is a phase-separated(immiscible) configuration in which the species 
with weaker intraspecies repulsive interaction is surrounded by the other.
In the phase-separated domain, the interface of the TBEC is a circle when
the quasi-two dimensional trap is perfectly symmetric.
To initiate RTI, we decrease gradually the s-wave scattering length of the
outer species through a magnetic Feshbach resonance. As RTI sets in, the
outer species tends to sink to the center of the trap and instabilities begin 
to occur on the circular interface separating the two components. Now, if the
anisotropy of the trap is increased along a particular direction, the
circular interface evolves into an elliptic cylindrical one. Due to RTI, the
nature of various non-linear patterns developed on the interface changes on
varying the geometry of the trapping potential. It has been observed that
at a critical value of the anisotropy parameter, the normal modes on the 
interface bifurcates.\\
This paper is organized as follows: In Section \ref{pancake},  we formulate
the problem using mean-field dynamics in a quasi-two dimensional harmonic
trap. In Section \ref{geometry} \& \ref{geometry1}, 
we discuss about the interface geometry and normal modes of the interface and 
formulate the Helmholtz equation using elliptic cylindrical coordinates. In
Section \ref{geometry2}, we derive an analytic condition for the temporal
decay constant in an elliptic cylindrical interface. Lastly, in Section
\ref{numerical}, we present numerical results showing the dynamics of TBEC
as a result of RTI.


\section{Phase separated pancake shaped TBECs}
\label{pancake}
In the mean field approximation, the TBEC is described by a set of coupled 
Gross-Pitaevskii equations
\begin{equation}
  \left[ \frac{-\hbar^2}{2m_i}\nabla^2 + V_i(x,y,z) + 
         \sum_{j=1}^2g_{ij}|\Psi_j|^2 \right]\Psi_{i} =
         i\hbar \frac{\partial \Psi_{j}}{\partial t} , 
  \label{eq.gp}
\end{equation}
where $i = 1, 2$ is the species index, $g_{ii} = 4\pi\hbar^2a_i/m_i$ 
with $m_i$ as mass and $a_i$ as $s$-wave scattering length, is the 
intra-species interaction; $g_{ij}=2\pi\hbar^2a_{ij}/m_{ij}$ with 
$m_{ij}=m_i m_j/(m_i+m_j)$ as reduced mass and $a_{ij}$ as inter-species 
scattering length, is inter-species interaction and $\mu_i$ is the chemical 
potential of the $i$th species. The trapping potential is 
\begin{equation}
  V_{i}(x,y,z) = \frac{m_{i}\omega^{2}}{2}(x^{2} + \alpha_{i}^{2}y^{2} + 
  \lambda_{i}^{2}z^{2})
\label{trap_3d}
\end{equation}
where, $\omega$ is the radial trap frequency, considered identical, for the two 
components, and $\alpha_{i}, \lambda_{i}$ are the anisotropy parameters. For 
simplicity of analysis, we consider trap potentials of both the species 
have the same geometry $\alpha_{1} = \alpha_{2} = \alpha$ , 
$\lambda_{1} = \lambda_{2} = \lambda$ and $m_{1} = m_{2} = m$. The energy of 
the TBEC is 
\begin{subequations}
 \begin{align}
 E = &\,\int_{-\infty}^{\infty}\biggr[\sum_{i=1}^{2}\biggr(\frac{\hbar^{2}}{2m}
    |\nabla\Psi_{i}|^{2} + V_{i}(x,y,z)\Psi_{i}^{2} \notag\\
  &\, +\frac{U_{ii}}{2}|\Psi_{i}|^{4}\biggr)+U_{12}|\Psi_{1}|^{2}|\Psi_{2}|^{2}
   \biggr]dx\,dy\,dz
 \end{align}
\label{eq.gp3}
\end{subequations}
To express the energy in suitable units, we define the oscillator length
of the trapping potential $ a_{\rm osc}= \sqrt{\hbar/ (m\omega)}$
and consider $\hbar\omega$ as the unit of energy. We then divide
Eq.(\ref{eq.gp3}) by $\hbar\omega$ and apply the transformations
$ \tilde{x} = x/a_{\rm osc}$ , $\tilde{y} = y/a_{\rm osc}$, 
$\tilde{z} = z/a_{\rm osc}$, $\tilde{t} = t\omega$,  and 
$\tilde{E} = E/(\hbar\omega)$. The transformed order parameter
\begin{equation}
\Phi_{i}(\tilde{x},\tilde{y},\tilde{z}) =
\sqrt{\frac{a_{\rm osc}^{3}}{N_{i}}}\Psi_{i}(x,y,z)
\end{equation}
and energy of TBEC in scaled units is given by
\begin{subequations}
\begin{align}
\tilde{E} =
&\,\int\,d\tilde{x}d\tilde{y}d\tilde{z}\biggr(
    \sum_{i=1}^{2}N_{i}\biggr[\frac{1}{2}|\nabla\Phi_{i}|^{2}+
     V_{i}(\tilde{x},\tilde{y},\tilde{z})|\Phi_{i}|^{2}\notag\\
&\,+ N_{i}\frac{\tilde{U}_{ii}}{2}|\Phi_{i}|^{4}\biggr]+N_{1}N_{2}\tilde{U}_{12}   |\Phi_{1}|^{2}|\Phi_{2}|^{2}\biggr)
\end{align}
\end{subequations}
where, $\tilde{U}_{ii} = 4\pi a_{ii}/a_{osc}$ and 
 $\tilde{U}_{12} = 4\pi a_{12}/a_{osc}$. For simplicity of notations,
from here on we will represent the transformed quantities without tilde. 
Thus, in scaled units, the coupled 3D GP equation is given by
\begin{equation}
 \biggr[-\nabla^{2} + V_{i}(x,y,z) +
 \sum_{j=1}^{2}G_{ij}|\Phi_{j}|^{2}\biggr]\Phi_{i} = \mu_{i}\Phi_{i}
\label{eq.gp4}
\end{equation}
where, $G_{ii} = N_{i}\tilde{U}_{ii}$ and $G_{ij} = N_{j}\tilde{U}_{ij}$.
For the present work, we consider a pancake shaped trap, the 
axial frequency is much larger than the radial frequency($\lambda\gg1$).  
In this situation, the transformed order parameter $\Phi(x,y,z)$ is
factorized into
\begin{equation}
  \Phi(x,y,z) = \phi(x,y)\zeta(z)
\label{eq.gp5}
\end{equation}
where, $\zeta(z)$ is the normalized state of axial trapping potential 
$V_{i}^{axial} = \lambda^2 z^{2}/2$. From Eq.(\ref{eq.gp4}) after integrating 
out the axial order parameter, we obtain the scaled coupled 2D GP equations
\begin{equation}
  \biggr[-\nabla_{\perp}^{2}+ V_{i}(x,y) + \sum_{j=1}^{2}
  \mathcal{N}_{ij}|\phi_{j}(x,y)|^{2}\biggr]\phi_{i}(x,y) = \mu_{i}\phi_{i}(x,y)
\label{eq.gp6}
\end{equation}
where, $\nabla_{\perp}^2 = \partial_x^2 + \partial_y^2$, 
$\mathcal{N}_{ii} = 4 N_{i}\sqrt{2\pi\lambda}\,a_{ii}$ and 
$\mathcal{N}_{ij} = 4 N_{j}\sqrt{2\pi\lambda}\,a_{ij}$.  Using Thomas-Fermi 
approximation in  Eq.(\ref{eq.gp6}), 
one can show the two components are phase-separated when 
$({\mathcal N}_{12}>\sqrt{{\mathcal N}_{11}{\mathcal N}_{22}})$,
where, ${\mathcal N}_{ii}$ and ${\mathcal N}_{ij}$ are all positive. To examine RTI, we consider the phase separated state in 
axis symmetric trapping potentials with coincident centers and numerically 
solve the pair of time-dependent GP equations
\begin{equation}
  \biggr[-\nabla_{\perp}^{2}+ V_{i}(x,y) + \sum_{j=1}^{2}
    \mathcal{N}_{ij}|\phi_{j}(x,y)|^{2}\biggr]\phi_{i}(x,y) = i\frac{\partial
    \phi_{i}}{\partial t},
\label{eq.gp6a}
\end{equation}
to study the dynamical evolution. 

\begin{figure}[h]
\includegraphics[width=8.5cm]{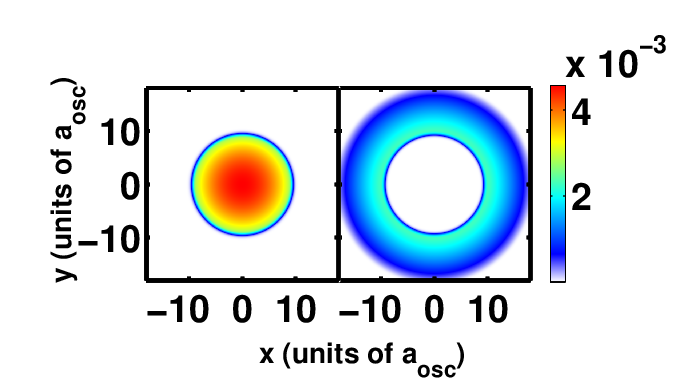}
\caption{Phase separated profiles of $^{85}$Rb--$^{87}$Rb mixture at
         $t=0$. The figure on the left shows the inner species($^{87}$Rb)
         with $a_{22} = 99a_{\rm B}$. The figure on the right shows the outer
          species($^{85}$Rb) with $a_{11} = 460 a_{\rm B}$.}
\label{int}
\end{figure}


\section{Interface geometry and modes}
\label{geometry} 
 In the phase-separated domain, the interface of the TBEC is a circle when
$\alpha $ is unity. It is, however, transformed to an ellipse when 
$\alpha >1$. A typical density profile of the the phase-separated TBEC
with $\alpha=1$ is shown in Fig. \ref{int}. Compared to Eq. (\ref{trap_3d}), a
more general form of 2D trapping potential is 
$V(x, y) = m(\omega_{x}^{2}x^{2} + \omega_{y}^{2}y^{2})/2$, where
$\omega_{x,y}$ represent angular trapping frequency along $x$ or $y$. 

Defining the geometric mean $\bar{\omega} = \sqrt{\omega_{x}\omega_{y}}$, the
trapping potential is 
\bea
V(x, y) = \frac{1}{2}m\bar{\omega}^{2}\left(\frac{x^{2}}{\beta^{2}} +
              \frac{y^{2}}{\gamma^{2}}\right ),
\label{eq.gp6c}
\eea
where, $\beta = \bar{\omega}/\omega_{x}= \sqrt{\alpha} $ and 
$\gamma =\bar{\omega}/\omega_{y}=1/\sqrt{\alpha}$. 
The density distribution of the TBEC, at moderate anisotropies, follows the 
geometry of the trapping potential. At larger anisotropies the interface 
energies modifies the density distribution and leads to difference from the 
geometry of the trapping potential. For the present study, we consider the TBEC
at moderate anisotropies. The interface of the TBEC is then an ellipse 
\begin{equation}
\frac{x^{2}}{\beta^{2}} + \frac{y^{2}}{\gamma^{2}} = 1,
\label{eq.gp6d}
\end{equation}
corresponding to the anisotropy parameters of the trapping potential. 

\begin{figure}[h]
\includegraphics[width=8.5cm]{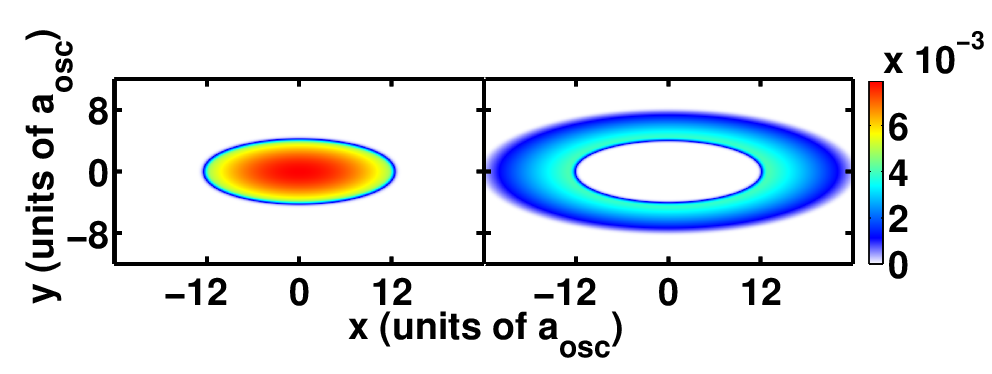}
\caption{Phase separated profiles of $^{85}$Rb--$^{87}$Rb mixture at
         $t=0$. The figure on the left shows the inner species($^{87}$Rb)
         with $a_{22} = 99a_{\rm B}$. The figure on the right shows the 
         outer species($^{85}$Rb) with $a_{11} = 460 a_{\rm B}$.}
\label{al3_int.png}
\end{figure}

At the interface, the densities are low, neglecting the intraspecies and 
interspecies 
interaction term, we get from Eq.(\ref{eq.gp6})
\begin{equation}
(-\nabla_{\perp}^{2} + V_{i})\phi_{i} = \tilde{\mu_{i}}\phi_{i}.
\label{eq.gp6e}
\end{equation}
where, $\tilde{\mu_{i}} = \mu_{i} - \sum_{j=1}^{2}
\mathcal{N}_{ij}|\phi_{j}(x,y)|^{2}$ .\\
Using Eq.(\ref{eq.gp6d}) and Eq.(\ref{eq.gp6e}), we get(in scaled units)
\begin{equation}
  \nabla_{\perp}^{2}\phi_{i} + \biggr(\tilde{\mu_{i}} - 
    \alpha\biggr) \phi_{i} = 0.
\label{eq.gp6f}        
\end{equation}
Defining the parameter $k_{i}^{2} = \tilde{\mu_{i}} - \alpha$,
the equation is 
\begin{equation}
(\nabla_{\perp}^{2} + k_{i}^{2})\phi_{i} = 0.
 \label{helm_2d}
\end{equation}
This is the Helmholtz equation in 2D. It must, however, be emphasized that the
equation is valid only at the interface or close to it. Away from the 
interface the densities are not small and intraspecies interactions is large.


\subsection{Normal modes of the interface}
\label{geometry1}

For linear stability analysis of the interface modes due to a small 
perturbation, to identify the onset of RTI, we transform the 
Eq. (\ref{helm_2d}) to elliptic cylindrical coordinates $(u, v)$. 
Here, the coordinate $v$ represent the asymptotic angle of confocal hyperbolic cylinders symmetrical about the x-axis. And, the $u$ coordinates are confocal 
elliptic cylinders centered on the origin\cite{gutierrez_03}. 
The transformation is defined by the relations $ x = a\cosh u\cos v$, 
and $ y = a\sinh u\sin v$ and $a$ is the focal distance along $x$-axis. 
We take the coordinates on the $z=0$ plane as we consider the TBEC in 2D. 
The Eq. (\ref{helm_2d}) then 
assumes the form
\begin{equation}
  \frac{1}{a^{2}({\rm sinh}^{2}\,u + {\rm sin}^{2}v)}
  \biggr(\frac{\partial^{2}\phi}{\partial u^{2}} 
 + \frac{\partial^{2}\phi}{\partial v^{2}}\biggr) + k^{2}\phi = 0,
   \label{eq.gp7}
\end{equation}
where $\phi$ is the solution of the form $\phi = U(u)\Theta(v)$.
Substituting, $\phi$ back in Eq.(\ref{eq.gp7}) we get,
\begin{equation}
 \biggr(\frac{1}{U}\frac{d^{2}U}{d u^{2}}+c^{2}{\rm
 sinh}^{2}u\biggr) + \biggr(\frac{1}{\Theta}\frac{d^{2}\Theta}{d
 v^{2}}+c^{2}{\rm sin}^{2}v\biggr) = 0.
 \label{eq.gp7a}
\end{equation}

Using separation of variables, the equation is simplified to the Mathieu 
equations \cite{mclachlan_51,gutierrez_03}
\begin{eqnarray}
\frac{d^{2}U}{d u^{2}} - \left ( {\mathcal A} - 2q\cosh 2u\right ) U=0, 
   \label{mathieu_1}  \\
\frac{d^{2}\Theta}{d v^{2}} + \left ( {\mathcal A} - 2q\cos 2v\right
)\Theta=0,
   \label{mathieu_2}
\end{eqnarray}
where,  ${\mathcal A} = A + a^2k^2/2$ and $q = a^2k^2/4 $. Here $A$ is the 
separation constant and returning to the earlier definition of the trapping 
potential, the anisotropy parameter $\alpha = \beta/\gamma $. The interface 
is an ellipse with eccentricity $ e = \sqrt{1 - 1/\alpha^{2}}$ and from the
theory of conic sections $a=\beta e = \sqrt{\alpha}e $. Based on these 
definitions, the constants in the Eq. (\ref{mathieu_1}) and (\ref{mathieu_2}) 
are redefined as
\begin{eqnarray}
   q & = &\frac{1}{4}k^2e^2\alpha , 
   \label{q_def}    \\
   {\mathcal A} & = & A+\frac{1}{2}k^2 e^2\alpha.
\end{eqnarray}
The constants in this form are easier to connect with the parameters of
trapping potentials. The Eqns.(\ref{mathieu_1} and \ref{mathieu_2}) then
assumes the form
\begin{eqnarray}
   \frac{d^{2}U}{d u^{2}} - \left [A + \frac{1}{2}k^2e^2\alpha
      \left (1 - \cosh 2u \right ) \right ]U &=& 0 ,
     \label{mathieu-1} \\
   \frac{d^{2}\Theta}{d v^{2}} + \left [A + \frac{1}{2}k^2e^2\alpha
      \left (1 - \cos 2v \right ) \right]\Theta & = &0.
     \label{mathieu-2} 
\end{eqnarray}
The interface of the TBEC, an ellipse, has fixed coordinate $u$ representing 
the elliptic cylinder. But the angle coordinate $v$ varies and lies in the 
domain $[0, 2\pi)$. Thus $\Theta$, solutions of the second equation, 
represent the normal modes of the interface. For circular interface,
$\alpha = 1 $ and $e = 0$, only $0 < A$ is physically admissible and
the solution of the equation is reduced to sinusoidal functions.


\subsection{Instability at the interface}
\label{geometry2}

 For TBEC in traps,  the gradient of the trapping potential is like the 
gravitational force in the conventional fluid dynamics and the flows within
TBEC is  modelled as potential flows. Consider the interface of the TBEC,
using the method of normal modes, any arbitrary disturbance on the interface 
may be resolved into independent modes of the form
\begin{eqnarray}
\label{zeta}
\xi & = & \hat{\xi}\,\Theta(v)e^{st},\\
\label{phi1}
\phi_{1}^{'} & = & \hat{\phi_{1}}(u)\,\Theta(v)e^{st},\\ 
\label{phi2}
\phi_{2}^{'} & = & \hat{\phi_{2}}(u)\,\Theta(v)e^{st}.
\end{eqnarray}
Here, $\xi$ is the position of the interface relative to the equilibrium 
configuration, and $\phi_i^{'}$ is the increments in the velocity potential of 
the $i$th species about the interfacial region caused due to disturbance in 
the system.  $\hat{\xi}$ and $\hat{\phi_i}$ are the amplitude of the 
modes and $s$ is the temporal decay constant.

We know that for an incompressible fluid, the Laplacian of the velocity 
potential vanishes $ \nabla^{2}\phi_i^{'} = 0 $ and from the expression of the 
normal modes
\begin{equation}
  \frac{\partial^{2}\hat{\phi_i}}{\partial u^{2}} + \biggr(\frac{1}{\Theta}
  \frac{\partial^{2}\Theta}{\partial v^{2}}\biggr)\hat{\phi_i} = 0.
\label{sv1}
\end{equation}
Using separation of variables, Eq.(\ref{sv1}) can be simplified to, 
\begin{eqnarray}
\frac{\partial^{2}\hat{\phi_i}}{\partial u^{2}} - C^{2}\hat{\phi_i} = 0.
\label{1}
\end{eqnarray}
where, $C^{2}= -(1/\Theta)(\partial^{2}\Theta/\partial v^{2})$.
The general solution of the above equation in the regions of the two species
are 
\begin{eqnarray}
\hat{\phi_{1}}(u) = A_{1}e^{-Cu} + B_{1}e^{Cu}, \nonumber\\
\hat{\phi_{2}}(u) = B_{2}e^{-Cu} + A_{2}e^{Cu},
\end{eqnarray}
where, $A_{1}$, $B_{1}$, $A_{2}$, $B_{2}$ are arbitrary constants. Here, it is
to be noted that the sign of the exponents in the two solutions are 
interchanged. This is to indicate that the relative distance from the 
interface, within the two species, are in opposite directions. We may recall 
that the instabilities occur only at the interface or close to it. At any point
far removed from the interface the normal modes must decay to zero. 
Thus, normal modes are of the form $ \hat{\phi_{1}} = A_{1}\exp(-Cu) $ and 
$\hat{\phi_{2}} = A_{2}\exp(Cu)$. The velocity potentials in Eq.(\ref{phi1}) 
and Eq.(\ref{phi2}) are then
\begin{eqnarray}
\phi_{1}^{'} &=& A_{1}e^{-Cu}\,\Theta e^{st}, \\
\phi_{2}^{'} &=& A_{2}e^{Cu}\,\Theta e^{st}.
\end{eqnarray}
The dynamical evolution of the interface is
described through a combination of the continuity equation, Euler's
equation and Bernoulli's theorem\cite{drazin_04,chandrasekhar_81}.
For stability analysis of the interface, we linearize these equations   
and neglect quadratic terms in $\phi_{1}^{'}$, $\phi_{2}^{'}$ and $\xi$. 
After linearization, we get
\begin{eqnarray}
\frac{\partial \phi_{i}^{'}}{\partial u} & = & \frac{\partial \xi}{\partial t},
\label{4}    \\
n_{1}\biggr(\frac{\partial \phi_{1}^{'}}{\partial t} + g\xi \biggr) & = &
n_{2}\biggr(\frac{\partial \phi_{2}^{'}}{\partial t} + g\xi \biggr),
\label{5}
\end{eqnarray}
where, $g$ is the gradient of the trapping potential $V(x,y)$. On the
interface, using the solutions obtained earlier, from Eq. (\ref{5}) one can 
show that 
\begin{equation}
n_{1}(s A_{1} + g \hat{\xi}) = n_{2}(s A_{2} + g \hat{\xi}).
\label{rln}
\end{equation}
In a similar way, from Eq. (\ref{4}), we obtain $A_{1} = -s\hat{\xi}/C$ and 
$A_{2} = s\hat{\xi}/C$. Using these values in the above equation
\begin{equation}
n_{1}\biggr(-\frac{s^{2}\hat{\xi}}{C} + g\hat{\xi}\biggr) = 
n_{2}\biggr(\frac{s^{2}\hat{\xi}}{C} + g\hat{\xi}\biggr).
\label{rln1}
\end{equation}
Simplifying this equation, one arrives at the definition of the
 temporal decay constant
\begin{eqnarray}
s = \pm\left [ \frac{Cg(n_{1} - n_{2})}{n_{1} + n_{2}}\right ]^{\frac{1}{2}}.
\end{eqnarray}
The densities of the condensates $n_{1}$ and $n_{2}$ are at a point
$(u,v)$ on the interface. We recollect that $n_{2}$ refer to the density of 
species at the center which is surrounded by the species with density
$n_{1}$. The interface is stable when $s$ is imaginary ( $n_{1}<n_{2}$) and
oscillates when perturbed. However, when $n_{1}>n_{2}$, the value of $s$ is
real and any perturbation, however small, grows exponentially with time.
This is the prerequisite for RTI in binary condensates. In this context,
Atwood number ${\Gamma }$ is given by \cite{sharp_84}
\begin{equation}
{\Gamma }=\frac{n_{1} - n_{2}}{n_{1} + n_{2}},
\end{equation}
>From Eq. (\ref{mathieu-2})
\begin{equation}
   C^2  =  \left [A + \frac{1}{2}k^2e^2\alpha \left (1 - \cos 2v\right )  
           \right],
\end{equation}
the temporal decay constant is then
\begin{equation}
s = \pm\left [A + 2q\left (1 - \cos 2v\right )  
    \right]^{\frac{1}{4}} \left [ \frac{g(n_{1} - n_{2})}{n_{1} 
    + n_{2}}\right ]^{\frac{1}{2}},
\label{tempd}    
\end{equation}
where for compact notation we have used the relation $q = k^2e^2\alpha/4$ 
given earlier. Thus Eq. \ref{tempd} can be rewritten as
\begin{equation}
s = \pm\left [A + 2q\left (1 - \cos 2v\right )  
    \right]^{\frac{1}{4}} \sqrt{{\Gamma } g}.
\end{equation}
This shows that $s$ is a function of $v$, the angular
coordinate.

\begin{figure}[h]
\includegraphics[width=7cm]{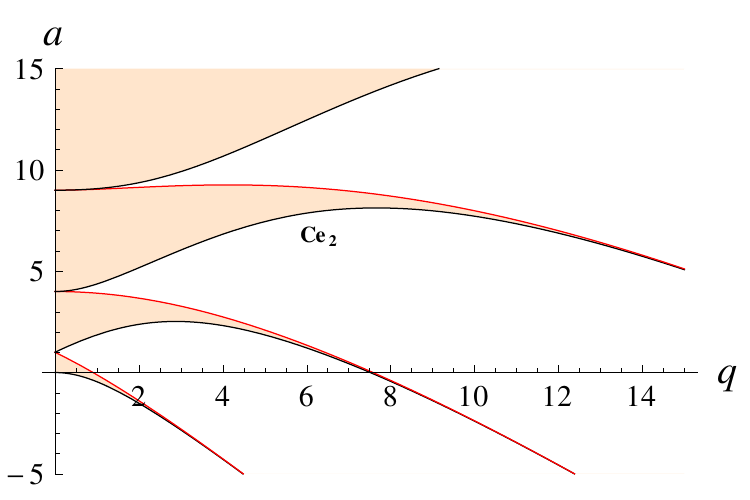}
\caption{Shaded regions indicate the values in $aq$-plane where solutions of 
         angular Mathieu equation, ${\rm ce}_{\nu}(v, q)$ and 
         $ {\rm se}_{\nu}(v, q)$, exist. The black (red) colored curves
         are the values of $a$ and $q$ for which integer order,
         ${\rm ce}_{n-1}(v, q)$ ($ {\rm se}_n(v, q)$ ) with 
         $n=1, 2, 3, \ldots$, solutions exist.
         }
\label{mathieu-aq}
\end{figure}


\subsection{Allowed solutions}

 The solutions of Eq. (\ref{mathieu-2}), the angular Mathieu equation, are
the ${\rm ce}_{\nu}(v, q)$ and $ {\rm se}_{\nu}(v, q)$ functions 
\cite{wolf-10}, cosine and sine elliptic functions, respectively. Here, $\nu$ 
is real number and denotes the order of the elliptic functions. The solutions, 
however, exist only for certain range of $a$ and $q$, and these are shown as 
shaded regions in Fig. \ref{mathieu-aq}. In the figure, the shaded region 
consists of lobes and each are bounded by elliptic function of integer orders 
${\rm ce}_{n-1}(v, q)$ and $ {\rm se}_n(v, q)$, where $ n=1, 2, 3, \ldots$.
For the present case, when RTI sets in, the mushroom shaped superfluid flows 
have four fold symmetry in the case of circular symmetry. So that the flow
retains symmetry or shape invariance along perpendicular directions. The 
corresponding solution of Eq. (\ref{mathieu-2}) which satisfy this condition 
is then ${\rm ce}_2(v, q)$, and it has the properties 
${\rm ce}_2(v, q)= {\rm ce}_2(v+\pi, q)$  and 
${\rm ce}_2(v, q)= {\rm ce}_2(-v, q)$. The loci of the $a$-$q$ pairings which 
allow this solution is the labeled curve in Fig. \ref{mathieu-aq}.

  One property of ${\rm ce}_2(v, q)$ is, the maximum at $v=0$ undergoes a 
smooth bifurcation at higher values of $q$.  Coming to the description of 
the interface in the binary condensates, from Eq. (\ref{q_def}), $q$ is a 
linear function of the anisotropy parameter $\alpha$. So, as we increase 
the anisotropy the mushroom shaped flows in RTI must undergo bifurcation.
At some value of $\alpha$, instead of four there must be six mushroom shaped 
inward superfluid flow.


\section{Numerical results}
\label{numerical}
To corroborate the analytic results for the interface modes,
as mentioned earlier, we numerically solve the pair of coupled
Eq. \ref{eq.gp6a}. We resort to split-step Crank-Nicholson
method\cite{muruganandam_09} implemented for binary condensates. 
We discretize Eq. \ref{eq.gp6a} both in space and time, and propagate the
resulting discretized equation in imaginary time, over small time steps.
In imaginary-time propagation method, $t$ in Eq. \ref{eq.gp6a} is replaced 
by $-i\tau$. This method seems to be more appropriate as the stationary 
ground state wave function of the TBEC is essentially real and dealing with 
real variables is more convenient than imaginary ones.
The split-step imaginary time solution obtained in a self-consistent way
after several iterations, is the stationary state of TBEC for the given
parameters used in this paper. The time-independent solution of TBEC, 
thus obtained, is used as an initial state for real-time propagation. 
The real-time propagation method yields the solution of time-dependent 
GP equation for TBEC, which is used to study the dynamical behaviour of
TBEC.

\begin{figure}[h]
 \includegraphics[width=8.5cm]{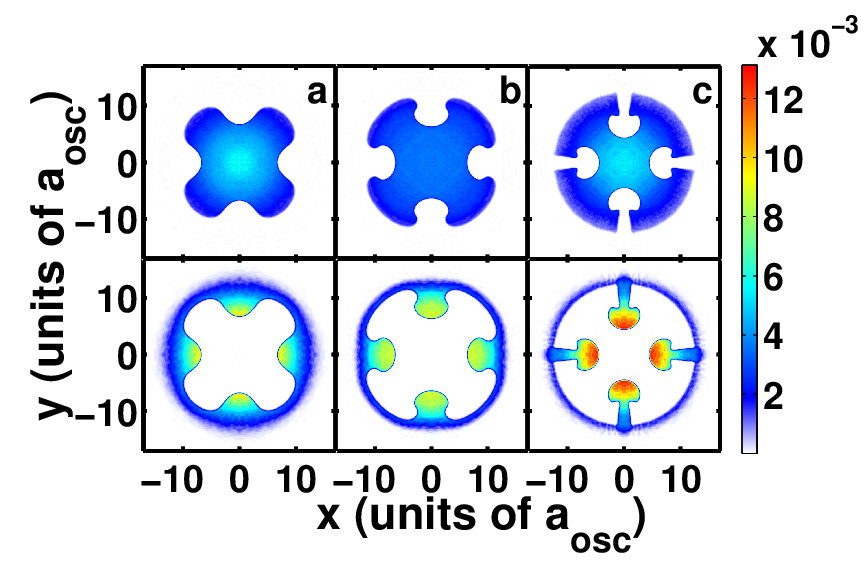}
 \caption{(a)-(c) shows development of mushroom shape pattern on the
          interface after $t = 358$ ms, $t = 378$ ms, $t = 400$ ms. The 
          scattering length is decreased from $a_{11} = 460 a_{\rm B}$ to 
          $a_{11} = 55 a_{\rm B}$ between $t = 0$ ms and $t = 200$ ms , after 
          that $a_{11}$ is fixed to $55 a_{\rm B}$ upto $t = 400$ ms. The 
          images on the upper panel correspond to the inner species 
          ($^{87}$Rb) and the images on the lower panel correspond to the
          outer species ($^{85}$Rb).}
\label{mushroom}
\end{figure}

As a representative case, we numerically calculate the stationary state
solution of TBEC based on the aforementioned method, with the parameters given 
in Ref. \cite{papp_08}. To study RTI, we use the imaginary-time
solution as the initial state. With the propagation of this solution over
real-time, we gradually change the scattering length over time steps, and
study its dynamics \cite{gautam_10}. Density profiles as shown in 
Fig. \ref{mushroom}, are the numerical solutions obtained by this method.


\subsection{Mode bifurcation and density profiles}
\label{numerical1a}
We consider a system of $^{85}$Rb--$^{87}$Rb atoms in a symmetric 2-D harmonic 
trapping potential with ($\omega_{\perp},\omega_{z}$) = $2\pi \times (8,90)
{\rm Hz}$. We choose initial state
to be the ground state for which $a_{11} = 460 a_{\rm B}$, 
$a_{22} = 99a_{\rm B}$, $a_{12} = a_{21} = 214 a_{\rm B}$, with $a_{\rm B}$ 
being the Bohr radius. The number of atoms are $N_{1} = 5\times10^{5}$ and 
$N_{2} = 10^{6}$ \cite{papp_08}. In this configuration, 
component 1(outer), $^{85}$Rb  completely surrounds component 2(inner),
$^{87}$Rb. Fig. \ref{int} shows phase separated profiles of the TBEC at $t=0$ 
in a perfectly symmetric pancake shaped trap i.e. $\alpha = 1$.

Now, the $s$-wave scattering length $a_{11}$ of the outer species is decreased
gradually over time, experimentally this is possible through the 
$^{85}$Rb-$^{85}$Rb magnetic Feshbach resonance\cite{roberts_2000}. However, 
throughout the process, we maintain 
$({\mathcal N}_{12}>\sqrt{{\mathcal N}_{11}{\mathcal N}_{22}})$ so
that the TBEC remains in the immiscible domain. A stage is reached when
$a_{11} < a_{22}$, $^{85}$Rb-$^{85}$Rb interaction weaker than the
$^{87}$Rb-$^{87}$Rb interaction. In this situation, the existing spatial 
structure of the system is energetically unfavourable and the outer species 
starts penetrating inside the inner species. Instabilities begin to occur at
the interface of the two components and eventually grows into a four 
fold mushroom shape superfluid flow as shown in Fig. \ref{mushroom}
\begin{figure}[h]
 \includegraphics[width=8.5cm]{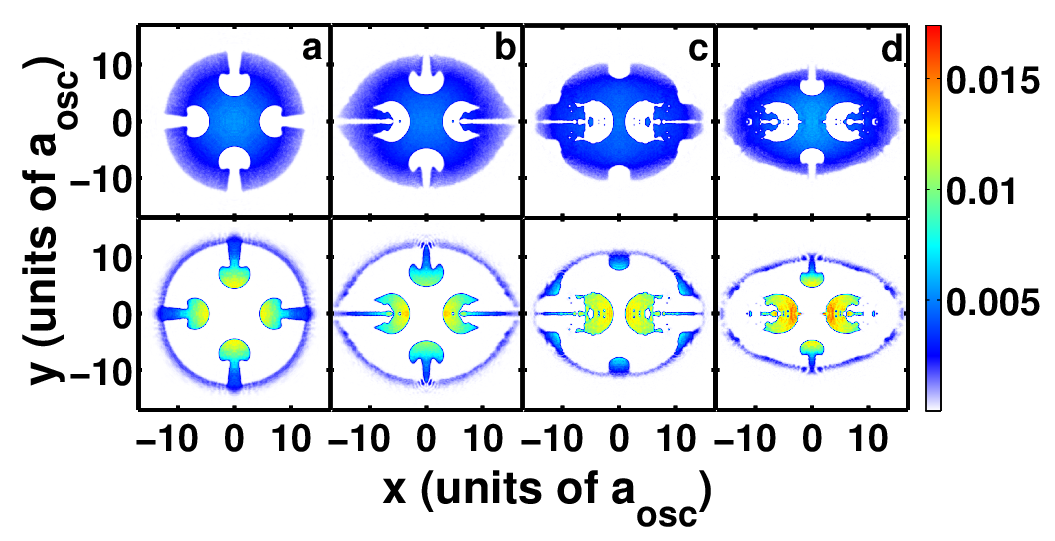}
 \caption{(a)-(d) Development of various non linear patterns for $\alpha = 1$,
          $\alpha = 1.2$, $\alpha = 1.4$ and $\alpha = 1.6$.The images on
          the upper panel correspond to the inner species ($^{87}$Rb) and 
          the images on the lower panel correspond to the
          outer species ($^{85}$Rb)}
\label{nonl1}
\end{figure}

\begin{figure}[h]
\includegraphics[width=8.5cm]{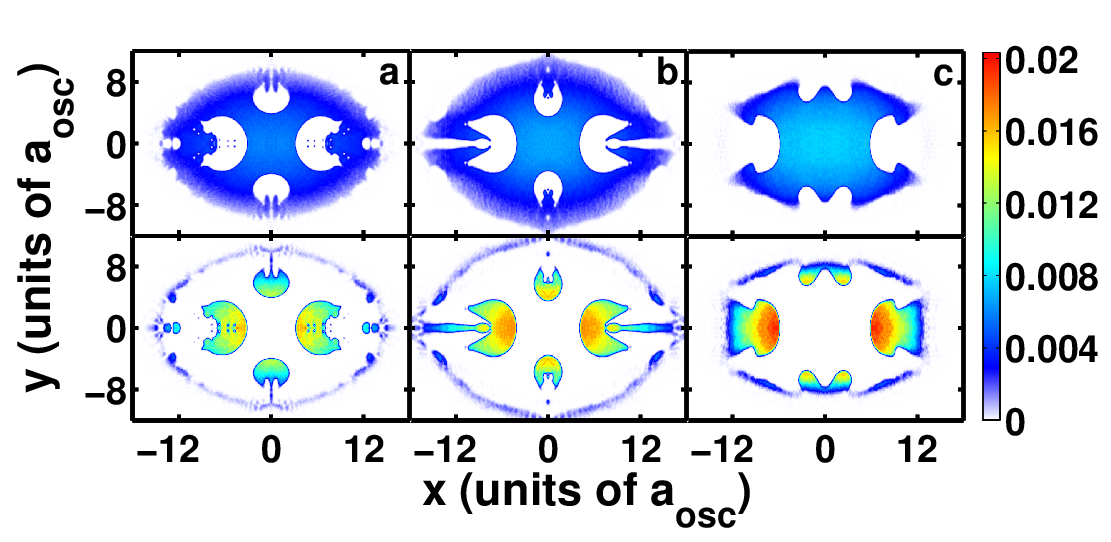}
\caption{(a)-(c) Development of various non linear patterns for $\alpha =
        1.8$, $\alpha = 2.0$ \& $\alpha = 3.0$. The images on the upper panel 
        correspond to the inner species ($^{87}$Rb) and the images on the 
        lower panel correspond to the outer species ($^{85}$Rb)}
                                      
\label{nonl2}
\end{figure}

The dynamics and the formation of lobes also depends on the geometry of the
interface. As the anisotropy of the trap $\alpha$ is increased keeping 
$\lambda$ and other remaining parameters fixed, the circular interface evolves 
into an elliptic interface. The penetration of the heavier fluid into the
lighter fluid gets initiated along the $x$-axis, followed by the
formation of lobes. This happens because, the interface is more curved along 
this direction with less confinement. Larger is the curvature, higher is the 
rate of inflow of the heavier fluid. Mass transport gradually occurs along 
$y$-direction, which is tightly confined. The interface here, is relatively 
flat and the lobes are formed at later stages of evolution. This is clearly
evident from the superfluid flow pattern soon after the onset of RTI as
shown in Fig. \ref{nonl1}. In Fig. \ref{nonl1}c-d, the lobes of $n_2$ along 
the $x$-axis are well developed and located deep within $n_1$. 
As $\alpha$ is increased further, the interface is deformed further. The
lobe along the $y$-axis undergoes a bifurcation when the anisotropy is such
that $ \alpha=3$ and the density profile of the superfluid flow is shown in 
Fig. \ref{nonl2}. Taking an average along the interface and close to the bulk
of $n_1$, when the mode bifurcates the value of $\tilde{\mu}$ is 
5.25. This can be related qualitatively to the analytic results, in which
case ${\rm ce}_2(v, q)$ undergoes bifurcation at around $q\approx 3.7$.  Thus
our numerical results is in agreement with the inferences drawn from 
the analytic solutions of the interface modes. 


\subsection{Bogoliubov analysis}
For a more detailed understanding of the instability, we perform a
Bogoliubov analysis for TBEC in a 2-D harmonic trap. Setting, $\Phi_{i} =
\phi_{i} + \delta\phi_{i}$, we expand the set of
coupled GP equations in Eq. \ref{eq.gp6a} in  $\delta \phi_{i}(x,y)$. Here,
$\phi_{i}$'s fixes the condensate density through $n_{i}(x,y) =
|\phi_{i}(x,y)|^{2}$ and $\delta \phi_{i}$'s are the deviations from the
initial ground state, which includes the quasi-particle excitations. We
consider excitation mode of the form 
\begin{equation}
\delta\phi_{i} = e^{-i\mu_{i} t/\hbar}[u_{i}e^{-i\omega t} -  
v_{i}^{*}e^{i\omega t}],
\end{equation}
where, $\mu_{i}$ is the chemical potential, $\omega$ is the excitation
frequency, and $u_{i}$ and $ v_{i}$ are the Bogoliubov amplitudes.
Using this ansatz, the Bogoliubov equations are
\begin{eqnarray}
  \biggr[-\frac{\hbar^{2}}{2m_{i}}\nabla_{\perp}^{2} + V_{i} & + 2 n_{i}U_{ii} 
  + U_{ij}n_{j} - \mu_{i} \biggr]u_{i} - U_{ii}n_{i}v_{i}  \nonumber\\
& + U_{ij}\sqrt{n_{i}n_{j}}(u_{j}-v_{j}) = \hbar\omega u_{i}, \\
\biggr[-\frac{\hbar^{2}}{2m_{i}}\nabla_{\perp}^{2} + V_{i} & +  2
n_{i}U_{ii} + U_{ij}n_{j} - \mu_{i} \biggr]v_{i} - U_{ii}n_{i}u_{i}  \nonumber\\
& + U_{ij}\sqrt{n_{i}n_{j}}(v_{j}-u_{j}) = -\hbar\omega v_{i}.
\end{eqnarray}
These equations are then numerically diagonalized to calculate the
excitation spectrum. If the frequencies are real, the perturbations 
remain bounded and the system is dynamically stable. On the other hand, 
pure imaginary eigenfrequencies denote instability of the system. The eigenmode 
corresponding to this complex frequency grows exponentially and is a signature 
of dynamically unstable system \cite{blaizot_86}.

  As a case study, we choose $N_{2}/N_{1} = 2$. For the isotropic case, 
$\alpha = 1$, we expand the Bogoliubov amplitudes in harmonic oscillator basis 
wave function. When $\delta \phi_{i}$ is small, Re$(\omega)$ increases 
monotonically upto a critical point as the $a_{11}$ is decreased. When 
$ a_{11}<a_{22}$ and RTI sets in, the low lying excitation modes $\omega$ 
starts becoming imaginary. The value of the Im$(\omega)$ increases 
monotonically as $a_{11}$ is decreased further and away from the critical 
point. These imaginary modes are signatures of instability in the dynamics
of the binary condensates.


\subsection{Effect of noise}
Numerical studies that have been carried out so far are at zero temperature
and without any imperfections, hence quite ideal. But in experiments, 
conditions are far from ideal. Fluctuations play a major role, and if large, 
may destroy the observed signatures predicted from the numerical simulations. 
One immediate remedy is to include fluctuation to our calculations. We 
introduce white noise during the real time evolution of TBEC. The white noise 
is at the level of $0.01\%$. Even after introducing noise, we still observe 
signatures of RTI as a result of changing $a_{11} $. The thermodynamical
quantities such as energy, chemical potential may vary quantitatively, but,
there is no qualitative difference in the shape of the interface after RTI
is initiated. Bifurcation of normal modes on the interface are still
observed at the predicted values of the anisotropy of the trap.


\section{conclusions}

  We have examined RTI at the interface of binary condensates as a function
of anisotropy parameter and ratio of number of atoms. The mushroom shaped
superfluid flow is four lobed, as expected, at low anisotropies. Based on
the analytical studies, the lowest natural mode is ${\rm ce}_{n-1}(v, q)$,
which describes the four lobed superfluid flow. However, at higher 
anisotropies corresponding to larger values of $q$, one of the maxima
of ${\rm ce}_2(v, q)$ bifurcates. This is also observed in the 
numerical simulation of the RTI at higher $ \alpha$. The RTI and 
bifurcation of the mode is robust, and observable in presence of white
noise.

\begin{acknowledgments}
We thank S. Chattopadhyay and K. Suthar for useful discussions. 
The results presented in the paper are based on the computations using the 
3TFLOP HPC Cluster at Physical Research Laboratory, Ahmedabad.
\end{acknowledgments}

\bibliography{bec}
\bibliographystyle{apsrev4-1}
\end{document}